\documentclass[aps, prd, showpacs, showkeys, preprint]{revtex4}
\begin{document}
\begin{titlepage}
~\vskip1cm
\title{ Infrared freezing of Euclidean observables and analyticity in perturbative QCD}
\author{Irinel Caprini\footnote{caprini@theory.nipne.ro}}
\affiliation{National Institute of Physics and Nuclear Engineering,Bucharest POB MG-6, R-077125 Romania}
\author{Jan Fischer\footnote{fischer@fzu.cz}}
\affiliation{Institute of Physics, Academy of Sciences of the Czech Republic,CZ-182 21  Prague 8, Czech Republic}

\begin{abstract}   The  renormalization-group improved finite order expansions of the QCD observables 
have an unphysical singularity in the Euclidean region, due to the Landau pole of the running coupling. 
Recently it was claimed that, by using a modified Borel representation, the leading one-chain term in 
a skeleton expansion of the Euclidean QCD observables is finite and continuous across the Landau pole, 
and then exhibits an infrared freezing behaviour, vanishing at $Q^2=0$. In the present paper we show, 
using for illustration the Adler-${\cal D}$ function, that the above Borel prescription  violates the 
causality  properties expressed by energy-plane analyticity: the function ${\cal D}(Q^2)$ thus defined 
is  the boundary value of a piecewise analytic function  in the complex  plane,  instead of  being 
a standard analytic function. So, the price to be paid for the infrared freezing of Euclidean 
QCD observables is the loss of a fundamental property of local quantum field theory.
\end{abstract}
\maketitle
 \end{titlepage} 
\newpage\section{Introduction}

The renormalization-group improved  expansion of the QCD physical amplitudes and Green 
functions is plagued at finite orders  by the unphysical singularity of the running 
coupling (the Landau pole). For instance, the expansion of the Euclidean Adler function 
in massless QCD 
\begin{equation}\label{Dseries} 
{\cal D}^{(N)}_{PT} (Q^2) = \sum \limits_{n= 0}^N d_n\, a^{n+1}(Q^2)
\end{equation}
has  a pole on the spacelike axis at $Q^2=\Lambda^2$, present  in the one-loop coupling 
\begin{equation}\label{asrng}
   a(Q^2) = \frac{\alpha_s(Q^2)}{\pi}= \frac{1}{\beta_0\ln(Q^2/\Lambda^2)} \,.
\end{equation} 
Here $\beta_0=(33- 2 n_f)/12$ is the leading beta-function coefficient in  QCD with $N_c=3$ and
$n_f$ active flavours, and $Q^2$ is the energy variable defined such as $Q^2>0$ on the 
spacelike (Euclidean) axis, and $Q^2<0$ on the timelike (Minkowskian) one.
 
At finite orders, a modification of perturbative QCD  leading to regular Euclidean 
observables  has been made in the so-called "analytic perturbation theory" 
\cite{Sh, ShSo}. The basic ingredient of this approach is the K\"allen-Lehmann 
representation of the Green functions in terms of their spectral densities. This restores 
the required analyticity at finite orders of the perturbation expansion: the powers 
$a^{n}(Q^2)$ of the  Euclidean coupling are replaced by a set of functions ${\cal A}_n(Q^2)$ 
which have no unphysical singularities at $Q^2>0$.

Beyond finite orders, the observables can be defined by a summation of the Borel
type. The Borel transform $B(u)$ of the Adler function has singularities on the 
real axis of the $u$-plane \cite{Beneke}: the ultraviolet (UV) renormalons  
along the range $u\leq -1$, and the infrared (IR) renormalons along $u \geq  2$ 
(we adopt the definition of  Borel transform used in \cite{MaNe}). While the 
Borel transform is, for a wide class of functions, uniquely determined once all 
the perturbation expansion coefficients are explicitly given, the determination 
of the function having a given perturbative (asymptotic) expansion is, on the 
other hand, infinitely ambiguous; not only due to the singularities, but 
because the contour of the Borel-type integral can be varied rather freely,  
without affecting the values of the expansion coefficients of the 
perturbation series.   

  In Ref. \cite{BrMa} the authors use, in two different parts of the Euclidean
region, two different summation prescriptions of one single perturbation series.
They do so by choosing two different integration contours of the Borel-Laplace 
integral, depending on the sign of the running coupling. For positive coupling,  
$a(Q^2)>0$, they choose the contour along the positive (IR renormalon) axis, 
\begin{equation}\label{Borelp}   
{\cal D}_{PT} (Q^2)=\frac{1}{\beta_0}
\int\limits_0^\infty\!\mbox{e}^{-u/(\beta_0 a(Q^2))} \, B(u)\,{\rm d}u, \quad\quad a(Q^2)>0,
\end{equation}
 while for negative coupling the integral is taken instead along the negative 
 (UV renormalon) axis:
\begin{equation}\label{Boreln}   
{\cal D}_{PT}(Q^2)=\frac{1}{\beta_0}
\int\limits_0^{-\infty}\!\mbox{e}^{-u/ (\beta_0 a(Q^2))} \, B(u)\,{\rm d}u,\quad \quad a(Q^2)<0.
\end{equation}
This dual definition of one physical quantity has been introduced in 
\cite{BrMa} because the integral (\ref{Borelp}) is divergent for $a(Q^2)<0$ 
and vice versa, the integral (\ref{Boreln}) is divergent for $a(Q^2)>0$. In 
the present paper we shall examine some consequences of this definition. 

As argued in \cite{BrMa}, this result can be expressed also in terms of the 
characteristic function $\omega_{\cal D}(\tau)$ defined by Neubert, by using the
method developed in Ref. \cite{MaNe}: 
\begin{equation}\label{Domega}
   {\cal D}_{PT}(Q^2) =  \int\limits_0^\infty\!   {\rm d}\tau\,\omega_{\cal D}(\tau)  a(\tau Q^2). \end{equation}
Of course, neither of the above integrals
 is well-defined: in (\ref{Borelp}) and (\ref{Boreln}) the integration path runs along the infrared and ultraviolet renormalons, respectively. In (\ref{Domega}), the integrand exhibits the singularity in the coupling $a(\tau Q^2)$ at $\tau=\Lambda^2/Q^2$. Regulating with the Principal Value and taking into account the continuity of the  characteristic function $\omega_{\cal D}(\tau)$  at $\tau=1$ \cite{MaNe}, the authors of \cite{BrMa} conclude from (\ref{Domega}) that the contribution of the leading chain of the skeleton expansion of  the Adler function (or other similar  Euclidean observables)
 is finite along the whole spacelike  axis $Q^2>0$ and approaches a zero limit at $Q^2=0$. So,  the unphysical Landau pole present in finite order expansions disappeared from the all-order summation of a certain class of Feynman diagrams.  

Of course, since the perturbative series of QCD is ambiguous, it is possible  
in principle to implement a desired property by a suitable summation 
prescription, or by choosing a different integration contour. 

The question is 
however whether the change of the prescription does not violate other
 fundamental requirements. It turns out that this is the case for the 
 prescription chosen in \cite{BrMa}. The
 deficiency regards the analytic properties in the complex plane: in 
 \cite{BrMa} the authors calculate the leading term of the skeleton expansion 
 of the Adler function  ${\cal D}(Q^2)$ only on the Euclidean axis. In the 
 present paper we consider the properties of this function in the complex 
 energy plane. We use the fact that the characteristic function considered 
 in  \cite{BrMa} is also the inverse Mellin transform of the Borel function 
 $B$ \cite{MaNe}. Then, using the techniques developed in Refs. \cite{CaNe, 
 CaFi2005}, we calculate the Adler function in the complex energy plane, for 
 the Borel prescription adopted in \cite{BrMa}. The  aim is to  see whether 
 the prescription satisfies the analyticity properties in the complex energy 
 plane required by causality in field theories \cite{Oehme}.

In the next section we review a few facts about the characteristic function and 
the Mellin transform of the Borel function. In Section III we calculate the 
Adler function in the complex energy plane, using the definition for the 
Borel summation adopted in \cite{BrMa}, and in the final Section we discuss the analytic properties of this function. 

\section{Characteristic function and inverse Mellin transform}
 The characteristic  function $\omega_{\cal D}$ appearing in  Eq. (\ref{Domega}) was introduced  in \cite{MaNe}, where it was denoted by
 $\widehat w_D$ (the same notation was used in \cite{CaNe, CaFi2005}).
 In order to facilitate the comparison with Ref. \cite{BrMa} we adopt the notation $\omega_{\cal D} (\tau)$ used in this work. 
 As shown in \cite{MaNe}, the function $\omega_{\cal D}$ is the inverse Mellin transform \cite{Tit} of the Borel function $B$:
\begin{equation}\label{omega}   
\omega_{\cal D}(\tau) = \frac{1}{2\pi i} 
  \int\limits_{u_0-i\infty}^{u_0+i\infty}\!{\rm d}u\,   B(u)\,\tau^{u-1} \,. 
  \end{equation}
The inverse relation 
\begin{equation}\label{omegainv}   
B(u) = \int\limits_0^\infty\!{\rm d}\tau\, 
  \omega_{\cal D}(\tau)\,\tau^{-u} \,, 
  \end{equation} 
  defines the function $B(u)$ in a strip parallel to the
imaginary axis with $-1<\mbox{Re}\,u <2$, where it is assumed to be analytic.  The function $\omega_{\cal D}(\tau)$ was calculated in \cite{MaNe} and was re-derived recently in Ref. \cite{BrMa}
 in the large-$\beta_0$ approximation (which gives the contribution of the leading one-chain term in the skeleton expansion). Using (\ref{omega}), the calculation is based on residues theorem: for
 $\tau<1 $ the integration contour is closed on the right half-$u$-plane, and the result is the sum over the residues  of 
the infrared renormalons;   for
 $\tau>1 $ the integration contour is closed on the left half-$u$-plane, and the result  contains the residues
 of the  ultraviolet renormalons.  The residues of the IR and UV renormalons satisfy some symmetry properties \cite{BrMa}, but their contributions are not equal. Therefore  $\omega_{\cal D}(\tau)$  has different analytic 
expressions, depending on whether 
$\tau$ is less or greater than 1. 
Following Ref. \cite{BrMa}, we denote the two branches of $\omega_{\cal D}$  by $\omega_{\cal D}^{IR}$ and  $\omega_{\cal D}^{UV}$,
 respectively (in Refs. \cite{MaNe, CaNe, CaFi2005},  $\omega_{\cal D}^{IR}$ was denoted by $\widehat w_D^{(<)}$, and $\omega_{\cal D}^{UV}$ by $\widehat w_D^{(>)}$).  According to the above discussion, the deformation of the contour in (\ref{omega}) gives
\begin{equation}\label{omegaIR}   
\omega_{\cal D}^{IR}(\tau) = \frac{1}{2\pi i} \left[\,
  \int\limits_{{\cal C_+}}\!{\rm d}u\,   B(u)\,\tau^{u-1}- 
\int\limits_{{\cal C_-}}\!{\rm d}u\,   B(u)\,\tau^{u-1} \right] \,, 
  \end{equation}
 where ${\cal C_\pm}$  are two parallel lines going from $0$ to $+\infty$ 
slightly above and below the real positive axis, and
\begin{equation}\label{omegaUV}   
\omega_{\cal D}^{UV}(\tau) = \frac{1}{2\pi i} \left [\,
  \int\limits_{{\cal C_+'}}\!{\rm d}u\,   B(u)\,\tau^{u-1}- 
\int\limits_{{\cal C_-'}}\!{\rm d}u\,   B(u)\,\tau^{u-1}\,\right] \,, 
  \end{equation}
where ${\cal C_\pm'}$  are two parallel lines  going from $0$ to $-\infty$ 
slightly above and below the real negative axis.

The explicit expressions of  $\omega_{\cal D}^{IR}$ and $\omega_{\cal D}^{UV}$ in the large-$\beta_0$ approximation \cite{MaNe} are
\begin{eqnarray}\label{IRUV}   
\omega_{\cal D}^{IR}(\tau) &=&
 \frac{8}{3} \left\{ \tau\left(    \frac74 - \ln\tau \right) + (1+\tau)\Big[ L_2(-\tau) + \ln\tau  
  \ln(1+\tau) \Big] \right\} \,,\\  \omega_{\cal D}^{UV} (\tau) &=& \frac{8}{3} \left\{ 1 + \ln\tau  
  + \left( \frac34 + \frac12 \ln\tau \right) \frac{1}{\tau}    + (1+\tau)\Big[ L_2(-\tau^{-1}) -
 \ln\tau \ln(1+\tau^{-1}) \Big]    \right\} \,,\nonumber
 \end{eqnarray}
where $L_2(x)=-\int_0^x {{\rm d}t\over t}\ln(1-t)$ is the Euler dilogarithm.
As shown in \cite{MaNe}, the function $\omega_{\cal D}(\tau)$, defined by its two branches given in  (\ref{IRUV}), is continuous  at $\tau=1$
(as are also its first three derivatives). 

In the above relations the variable $\tau$ was real. However, from
  Eqs. (\ref{IRUV}) it is clear that both functions $\omega_{\cal D}^{IR}(\tau)$   and  $\omega_{\cal D}^{UV}(\tau)$  are  in fact analytic  in the $\tau$-complex plane cut along the real negative axis. This fact will be useful below.

\section{Adler function in the complex plane}
 
In this section we calculate the Adler function for complex values of the energy, adopting the choice of the Borel-Laplace integral made in \cite{BrMa}.
Following Ref. \cite{BrMa}
we work in the $V$-scheme, where all the exponential dependence in the Borel-Laplace integrals (\ref{Borelp}) and (\ref{Boreln}) is 
absorbed in the running coupling, and denote by $\Lambda_V^2$  the corresponding QCD scale parameter. Also, to facilitate the comparison with \cite{BrMa}, we denote the complex energy variable by $Q^2$ (connected by $Q^2=-s$ to the  $s$-channel energy variable used in \cite{CaNe, CaFi2005}, 
for which $s>0$ is the timelike axis).

Let us consider $Q^2$ complex, first such that $|Q^2|>\Lambda_V^2$. Since in this case $\mbox{Re}\, a(Q^2)>0$ we use the choice (\ref{Borelp}) of the Borel-Laplace integral. The integral is not well-defined, due to the IR renormalons along $u>2/\beta_0$.  As 
usual, we adopt  the principal value ($PV$) prescription, taking
\begin{equation}\label{pv}
{\cal D}_{PT}(Q^2)={1\over 2} [{\cal D}^{(+)}(Q^2)+ {\cal D}^{(-)}(Q^2)]\,,
\end{equation}
the quantities ${\cal D}^{(\pm)}(Q^2)$ being defined as
\begin{equation}\label{pm}   
{\cal D}^{(\pm)}(Q^2)=\frac{1} {\beta_0}
\int\limits_{{\cal C}_\pm}\!{\rm e}^{-u/(\beta_0 a(Q^2))} \, B(u)\,{\rm d}u\,,
\end{equation} where ${\cal C_\pm}$  are two parallel lines 
slightly above and below the real positive axis (these lines were introduced already in Eq.(\ref{omegaIR})).

Our aim is to express the integrals (\ref{pm}) in terms of the 
characteristic function $\omega_{\cal D}$  defined in the previous section. 
To reach this, we shall use the method developed in Ref. \cite{CaNe} some 
time ago and used in a similar context. To make our exposition selfcontained, 
we shall expound the method here in some detail. 

As a first step, we have to pass from the integrals along the contours 
${\cal C_\pm}$ to integrals along a line parallel to the imaginary axis, where 
the representation (\ref{omegainv}) is valid. This can be achieved by rotating 
the integration contour from the real to the imaginary axis, provided the 
contribution of the circles  at infinity is negligible. Let us first consider 
a point in the upper half of the energy plane, for which $Q^2=|Q^2|\,e^{i\phi}$ 
with a phase $0<\phi<\pi$. Taking $u={\cal R}\,e^{i\theta}$ on a large 
semi-circle of radius ${\cal R}$, the relevant exponential appearing 
in the integrals (\ref{pm}) is
\begin{equation}\label{expon}
   \exp\left\{ -{\cal R}  \left[\ln\left(\frac{|Q^2|}{\Lambda_V^2}\right)\cos\theta 
   -\phi \sin\theta \right] \right\} \,.
\end{equation}
For $|Q^2|>\Lambda_V^2$, the exponential is negligible at large ${\cal R}$ 
for $\cos\theta>0$ and $\sin\theta<0$, {\it i.e.} for the fourth quadrant of 
the complex $u$-plane. The integration contour 
defining $ {\cal D}^{(-)}(Q^2)$ can be rotated to the negative imaginary $u$-axis, where
the representation (\ref{omegainv}) is valid. This leads to the double
integral
\begin{equation}\label{double}
   {\cal D}^{(-)}(Q^2) = \frac{1} {\beta_0} \int\limits_0^{-i\infty}\!{\rm d}u
   \int\limits_0^\infty\!{\rm d}\tau\,\omega_{\cal D}(\tau) \exp\left[
   -u  \left( \ln\frac{\tau|Q^2|}{\Lambda_V^2} + i \phi \right)
   \right] \,.
\end{equation}
The order of integrations over $\tau$ and $u$ can be interchanged,
since for positive $\phi $ the integral over $u$ is convergent and can
be easily performed. Expressed  in terms of the complex variable $Q^2$, the result is
\begin{equation}\label{Dminus}
    {\cal D}^{(-)}(Q^2)  = \frac{1}{\beta_0} \int\limits_0^\infty\!{\rm d}\tau\,
   \frac{\omega_{\cal D}(\tau)}{\ln(\tau Q^2/\Lambda_V^2)}= \int\limits_0^\infty\!{\rm d}\tau\,
   \omega_{\cal D}(\tau) a(\tau Q^2) \,.
\end{equation}
Consider now the evaluation of the function $  {\cal D}^{(+)}(Q^2) $ given by 
the integral along the contour ${\cal C_+}$ above the real axis. 
Naively, one might think to rotate the integration contour to the 
positive imaginary axis without crossing any singularities. However, 
this rotation is not allowed, because along the corresponding 
quarter of a circle $\sin\theta>0$, and the exponent (\ref{expon}) does not 
vanish at infinity for $0<\phi$. The way out is to perform again a 
rotation to the negative imaginary $u$ axis, for which the 
contribution of the circle at infinity vanishes. But in this rotation 
the contour crosses the positive real axis, and hence picks up 
the contributions of the IR renormalon singularities located along 
this line. This can be evaluated by comparing the expression (\ref{omegaIR}) 
of the function $\omega_{\cal D}^{IR}(\tau)$ with the definition 
(\ref{pm}) of the functions ${\cal D}^{(\pm)} $: they are connected by the change of variable $\tau=\exp [-1/(\beta_0 a(Q^2))]$. It follows that ${\cal D}^{(+)} $ can 
  be expressed in terms of   ${\cal D}^{(-)}$ as
\begin{equation}\label{dif}
   {\cal D}^{(+)}  =  {\cal D}^{(-)} + \frac{2\pi i}{\beta_0} \,
   \frac{\Lambda_V^2}{Q^2}\, \omega_{\cal D}^{IR} (\Lambda_V^2/Q^2) \,.
\end{equation}
The relations (\ref{pv}), (\ref{Dminus})  and (\ref{dif}) completely specify the
function ${\cal D}_{PT}(Q^2)$ for $ |Q^2|>\Lambda_V^2$,  in the upper half plane $\mbox{Im}\, Q^2>0$ :
\begin{equation}
 {\cal D}_{PT}(Q^2) =   \int\limits_0^\infty\!   {\rm d}\tau\,\omega_{\cal D}(\tau) a(\tau Q^2)
   + \frac{i\pi}{\beta_0}\,  \frac{\Lambda_V^2}{Q^2} \,   \omega_{\cal D}^{IR}(\Lambda_V^2/Q^2)\,.
\end{equation}
Using the same method, the function ${\cal D}_{PT}(Q^2)$ can be calculated in the 
lower half of the energy plane, where $Q^2=|Q^2|\mbox{e}^{i\phi}$ with
$-\pi<\phi<0$. In this case, the integral along ${\cal C_+}$ can be 
calculated by rotating the contour up to the positive imaginary $u$ 
axis, while for the integration along ${\cal C_-}$ one must first pass 
across the real axis and then rotate towards the positive imaginary axis. Combining the results, we obtain the following
expression for the Adler function for complex $Q^2$ with $|Q^2|>\Lambda_V^2$:
\begin{equation}\label{Dup}
   {\cal D}_{PT}(Q^2) =   \int\limits_0^\infty\!   {\rm d}\tau\,\omega_{\cal D}(\tau) a(\tau Q^2)
   \pm \frac{i\pi}{\beta_0} \, \frac{\Lambda_V^2}{Q^2} \,   \omega_{\cal D}^{IR}(\Lambda_V^2/Q^2) \,,
\end{equation}
where the $\pm$ signs correspond to $\mbox{Im}\, Q^2>0$ and $\mbox{Im}\, Q^2<0$, respectively.
We recall that the first term in (\ref{Dup}) is given by the integration with respect to $u$, and the last term is produced by the 
residues of the infrared renormalons picked up by crossing the positive axis in the Borel plane. 

 Let us consider now $|Q^2|< \Lambda_V^2$, when $\mbox{Re}\, a(Q^2)<0$. Following \cite{BrMa} we use the definition  (\ref{Boreln}) of the Borel-Laplace integral along the negative axis. In this case the integral is not defined due to the UV renormalons. The Principal Value prescription will be given by (\ref{pv}), where now the integrals are
\begin{equation}\label{pm1}   
{\cal D}^{(\pm)}(Q^2)=\frac{1}{\beta_0}
\int\limits_{{\cal C'}_\pm}\!{\rm e}^{-u/(\beta_0 a(Q^2))} \, B(u)\,{\rm d}u\,,
\end{equation} where ${\cal C'_\pm}$  are  two parallel lines 
slightly above and below the negative axis (they were introduced already in Eq. (\ref{omegaUV})).

We apply then the same techniques as above, by rotating the contours ${\cal C'_\pm}$ towards the imaginary axis in the $u$ plane, in order to use the representation (\ref{omegainv}). If the exponential (\ref{expon}) decreases we can make the rotation. If not, we must first cross the real axis and perform the rotation. The calculations proceed exactly as before, with the difference that now one picks up the contribution of the UV renormalons, according to the relation (\ref{omegaUV}). This leads to the expression of the Adler function for $|Q^2|<\Lambda_V^2 $
\begin{equation}\label{Dlow}
   {\cal D}_{PT}(Q^2) =  \int\limits_0^\infty\!   {\rm d}\tau\,\omega_{\cal D}(\tau)  a(\tau Q^2) 
   \pm \frac{i\pi}{\beta_0}\,  \frac{\Lambda_V^2}{Q^2} \,
\omega_{\cal D}^{UV}(\Lambda_V^2/Q^2) \,,
\end{equation}
where the signs correspond to $\mbox{Im}\, Q^2>0$ and $\mbox{Im}\, Q^2<0$, respectively.

\section{Analyticity}
 We first recall that, in confined gauge theories like QCD,  causality and unitarity imply that the Green functions and the physical amplitudes are analytic functions in the complex energy variables, with singularities at the hadronic unitarity thresholds \cite{Oehme}. 
In particular, the Adler function $D(Q^2)$ should be real analytic  
in the complex $Q^2$ plane, cut along the negative real axis from the 
threshold  $- 4 m_\pi^2$ for hadron production  to $-\infty$. In perturbative massless QCD  the first unitarity branch point lies at $Q^2=0$, but otherwise the analyticity in the complex plane should be 
preserved. This property 
is implemented by the well-known K\"allen-Lehmann representation
\begin{equation}\label{KL}
 {\cal  D}(Q^2)=\frac{Q^2}{\pi} \int\limits_0^\infty \frac{{\cal R}(s)\, \mbox{d} s}{(s+Q^2)^2}\,,
\end{equation}
where ${\cal R}(s)$ is the perturbative part of the $R_{e^+e^-}$ ratio. In what follows we check whether the  function ${\cal D}_{PT}$ derived with the Borel prescription adopted in \cite{BrMa}  satisfies  this requirement.

In  the previous section we derived the expressions of ${\cal D}_{PT}(Q^2)$ in the upper and lower halves of the $Q^2$ plane.
 Since $\omega_{\cal D}(\tau)$ is a real analytic 
 function in the cut $\tau$-plane ({\it i.e.} 
 $\omega_{\cal D}(\tau^*)=\omega^*_{\cal D}(\tau)$), it follows from (\ref{Dup}) 
 and (\ref{Dlow}) that the values in the upper and lower halves of the 
 $Q^2$-plane are complex conjugate to each other: ${\cal D}_{PT} 
 ((Q^2)^*)={\cal D}_{PT}^*(Q^2)$.

 We compute now the limit of these expressions when $Q^2$ is approaching the 
 Euclidean axis. Consider first that $Q^2$  tends to the real positive axis 
 from above, in the region $|Q^2|>\Lambda_V^2$. In this case we use the 
 expression (\ref{Dup}). The integrand has a pole  at $\tau=\Lambda_V^2/Q^2$.  
 Writing explicitly the real and the imaginary part of this pole, when one adds 
 to $Q^2$  a small positive imaginary part, we obtain, for real $Q^2>\Lambda^2$:
\begin{equation}\label{Dup+}
   {\cal D}_{PT}(Q^2+i\epsilon) =  \mbox{Re}\left[ \int\limits_0^\infty\!   
   {\rm d}\tau\,\omega_{\cal D}(\tau)  a(\tau Q^2)\right] -
    \frac{i\pi}{\beta_0} \, \frac{\Lambda_V^2}{Q^2} \,  
[ (\omega_{\cal D}(\Lambda_V^2/Q^2)-\omega_{\cal D}^{IR}(\Lambda_V^2/Q^2)]\,.
\end{equation}
We notice now that for an argument equal to $\Lambda_V^2/Q^2<1$, the function 
 $\omega_{\cal D}$ coincides with $\omega_{\cal D}^{IR}$, so the last term in 
 (\ref{Dup+}) vanishes: the imaginary part of the singularity of the integral 
 is exactly compensated by the additional term appearing in (\ref{Dup}).

 For $Q^2<\Lambda_V^2$, we obtain  in the same way from (\ref{Dlow})
\begin{equation}\label{Dlow+}
   {\cal D}_{PT}(Q^2+i\epsilon)=
 \mbox{Re}\left[ \int\limits_0^\infty\!   {\rm d}\tau\,\omega_{\cal D}(\tau)  
 a(\tau Q^2)\right] -
    \frac{i\pi}{\beta_0} \, \frac{\Lambda_V^2}{Q^2}\,  
[ (\omega_{\cal D}(\Lambda_V^2/Q^2)-\omega_{\cal D}^{UV}(\Lambda_V^2/Q^2)]\,.
\end{equation}
Again the last term in this relation vanishes, since for an argument 
$\Lambda_V^2/Q^2>1$ the function  $\omega_{\cal D}$ is equal to $\omega_{\cal 
D}^{UV}$. Moreover, one can easily see that the expressions of 
${\cal D}(Q^2-i\epsilon)$, obtained for $Q^2$  approaching the Euclidean 
axis from the lower half plane, differ from (\ref{Dup+}) and (\ref{Dlow+}) 
only by the sign in front of the last term, which again vanishes. Thus, for 
all $Q^2>0$, the functions (\ref{Dup}) and (\ref{Dlow}) approach the same 
expression
\begin{equation}\label{D+-}
   {\cal D}_{PT}(Q^2\pm i\epsilon) =  \mbox{Re}\left[ \int\limits_0^\infty\!   {\rm d}\tau\,\omega_{\cal D}(\tau)  a(\tau Q^2)\right]\,.
\end{equation}
 This coincides with the PV regulated integral of the Cauchy type 
 (\ref{Domega}) which, as shown in \cite{BrMa}, is finite and satisfies the infrared freezing. Since $\omega_{\cal D}(\tau)$  is holomorphic (infinitely differentiable) for all $\tau>0$ except $\tau=1$, the right-hand side of (\ref{D+-}) has all derivatives at all points $Q^2>0$ except $Q^2=\Lambda^2$,  where only the first three derivatives exist \cite{BrMa}. This means that (\ref{Dup}) and (\ref{Dlow}) define in fact analytic functions in the regions $|Q^2|>\Lambda_V^2$ and $|Q^2|<\Lambda_V^2$, respectively.

 However, as seen from (\ref{IRUV}), $\omega_{\cal D}^{IR}(\tau)$ 
 and $\omega_{\cal D}^{UV}(\tau)$ are different analytic functions. The 
 expressions (\ref{Dup}) and (\ref{Dlow}) show that ${\cal D}_{PT} (Q^2)$ coincides with a certain analytic function in the region $|Q^2|>\Lambda_V^2$, and with another analytic function in the region $|Q^2|<\Lambda_V^2$. So, the Adler function derived with the Borel representation adopted in \cite{BrMa} is not analytic, but only piecewise analytic. This is in conflict with the principle of analyticity implemented by the K\"allen-Lehmann representation (\ref{KL}). 
 
This result implies that the infrared freezing of the Euclidean observables achieved in \cite{BrMa} has had a price. It has been possible only at the expense of analyticity, which is a guiding principle and fundamental property of field theory, and to which also the authors of \cite{BrMa} and \cite{HoMa} repeatedly refer. The loss is not only of an academic interest: the analytical continuation is the only technique to obtain the Minkowskian observables from the Euclidean ones. Indeed, the Minkowskian quantities like the $R_{e^{+}e^{-}}(s)$ ratio are derived from the Euclidean ones by analytical continuation. Since the Borel-Laplace summation, as defined in \cite{BrMa}, does not satisfy analyticity, one cannot find the Minkowskian quantities in a 
consistent way: more exactly, it is impossible to define a contour, situated entirely inside the analyticity domain, which connects the deep Euclidean region with the low-energy Minkowskian one. 

As already pointed out, there are many different functions having the same asymptotic expansion in powers of the coupling. The inconsistency of the approach of refs. \cite{BrMa,HoMa} is not so much in choosing a specific definition of the integration contour, but in changing it, 
for technical reasons, during the calculation of a physical quantity.
Every definition should be motivated physically, and so should be 
every variation of it.  

From the above calculations we see that the representation of the correlation functions in terms of the characteristic functions by using the Mellin transform technique is particularly suitable for examining the analytic properties of the correlation functions and performing their analytic continuation to $|Q^2|<\Lambda_V^2$ and to the Minkowskian region. We mention in this context the calculation performed in  Ref. \cite{CaNe}, where the function  ${\cal D}_{PT}(Q^2)$ was derived  for $|Q^2|>\Lambda_V^2$  by using the Borel prescription (\ref{Borelp}), and then the result was analytically continued to $|Q^2|<\Lambda_V^2$. As shown in \cite{CaNe}, the function  ${\cal D}_{PT}(Q^2)$ calculated in this way exhibits a cut along the real axis between $0$ and $\Lambda_V^2$. Except for this unphysical singularity, the function is analytic in the plane cut for $Q^2>0$. Therefore, one can define in a consistent way the Minkowskian quantity  ${\cal R}$ by analytic continuation from the deep Euclidean region. 

We applied this method  in \cite{CaFi2005}, where we investigated also the question of the infrared freezing along the timelike axis. We used the inverse Mellin transform representation of ${\cal D}_{PT}(Q^2)$ derived in \cite{CaNe} and calculated the Minkowskian quantity  ${\cal R}(s)$  by analytical continuation. We find that while the fixed-order approximants
 ${\cal R}^{(N)}(s) = \frac{1}{\pi}{\rm Im}\Pi^{(N)}(s+i\epsilon)-1$ tend to a finite value ${\cal R}^{(N)}(0) = 1/\beta_{0}$ in the infrared limit $s \to 0$ for all $N=1,2,3,...$ (thereby exhibiting infrared freezing), the corresponding Borel-summed all-orders quantity ${\cal R}(s)$ is divergent with $s \to 0$:
\begin{equation}
{\cal R}(s) = 1/\beta_{0} + \lim_{s \to 0}{\rm Re}\int_{-\Lambda^{2}_{V}/s}^{0}{\rm d}\tau {\hat w}^{(<)}_{D}(\tau).
\end{equation}
Indeed \cite{CaFi2005}, the additional integral on the right-hand side diverges like $\ln^{2}s /s^{2}$ for $s \to 0$. (Of course, one expects that in full QCD this divergent behavior will be compensated by a similar growth of similar terms in the OPE.)  

The above discussion refers only to the calculation in perturbation theory. In Ref. \cite{BrMa} the authors add to the perturbative Adler function a nonperturbative term. Although they use a slightly  different language, based on the ambiguity cancellation in the OPE, it can be seen from Eq. (81) of \cite{BrMa} that the nonperturbative part added to the perturbative function ${\cal D}_{PT}(Q^2)$ given in our relations (\ref{Dup}) and (\ref{Dlow})  has the form 
\begin{equation}\label{DNP}
   {\cal D}_{NP}(Q^2) = \kappa \, \frac{\Lambda_V^2}{Q^2} \, 
 \omega_{\cal D}(\Lambda_V^2/Q^2)\,,
\end{equation}
where $\kappa$ is a real constant.   Using the fact that   
$\omega_{\cal D}(\Lambda_V^2/Q^2)$ behaves at small $Q^2$ like 
$Q^4/\Lambda_V^4 \ln (\Lambda_V^2/Q^2)$  \cite{MaNe}, one can see 
from  the relations (\ref{D+-}) and (\ref{DNP}) that the  sum  
${\cal D}_{PT}(Q^2)+ {\cal D}_{NP}(Q^2)$ is finite along the  Euclidean 
axis and  vanishes at $Q^2=0$. But it fails to be a single analytic 
function in the complex $Q^2$-plane, being only piecewise analytic.

 In conclusion, we have shown by an explicit calculation that the Borel 
 prescription adopted in \cite{BrMa} is in conflict with the analyticity 
 requirements  derived from causality and unitarity in perturbative QCD.  
 Moreover, the simple model for the complete Adler function proposed in 
 \cite{BrMa} cannot represent the physical observable: although it is finite 
 in the Euclidean region and exhibits infrared freezing, it is not consistent 
 with the analyticity properties derived from field theory. 
 
\begin{acknowledgments} We acknowledge interesting discussions with Chris 
Maxwell on the subject of this work.  One of us (I.C.)  thanks Prof. 
Ji\v{r}\'{i} Ch\'{y}la  for hospitality at the Institute of Physics of the 
Czech Academy.  This work was  supported  by the CEEX Program of 
Romanian ANCS under Contract  Nr.2-CEx06-11-92, and by the Ministry of 
Education of  the Czech Republic, Project Nr. 1P04LA211. 
\end{acknowledgments}

\end{document}